\documentclass[twocolumn]{revtex4}

\usepackage{graphicx}

\usepackage{amsmath}
\usepackage{epsfig}

\begin{document}
 \title{Squeezing based on nondegenerate frequency doubling internal to a realistic laser}

\author{Ulrik L. Andersen\footnote[3]{Present address: Universit\"{a}t Erlangen-N\"{u}rnberg, Staudtstrasse 7/B2, D-91058 Erlangen, Germany. Electronic address: andersen@kerr.physik.uni-erlangen.de}}
\author{Peter Tidemand-Lichtenberg}
\author{Preben Buchhave}
\affiliation{Department of
 Physics, Technical University of Denmark, DK-2800 Kgs. Lyngby,
 Denmark.}
\begin{abstract}
We investigate theoretically the quantum fluctuations of the fundamental field
in the output of a nondegenerate second harmonic generation process
occuring inside a laser cavity. Due to the nondegenerate character of the
nonlinear medium, a field orthogonal to the laser field is for some operating
conditions indepedent of the fluctuations produced by the laser medium.
We show that this fact may lead to perfect squeezing for a certain polarization
mode of the fundamental field. The experimental feasibility of the system
is also discussed.
\end{abstract}


\maketitle

\section{Introduction}

In many quantum information applications a basic ingredient is a strongly squeezed Gaussian
state~\cite{furusawa98.sci,ralph00.pra}.
Unfortunately, the generation of such a state is often fraught with technical difficulties,
as demonstrated by vacuum squeezed light from a subthreshold optical
parametric oscillator system, which involves a laser, an upconversion process and the squeezing
interaction itself~\cite{breithenbach95.josab}. Owing to this
complexity it will appear fruitful to devise some alternative systems,
capable of producing efficient squeezing with a higher
degree of simplicity and hence scalability.
Towards this aim we investigate, theoretically, the amount of squeezing
that can be expected in a certain polarization state of a laser beam
using a polarization nondegenerate (type II) second harmonic
generation (SHG) process located inside a laser cavity.

The intensity noise of a standard laser operating well above laser threshold is usually at the
quantum noise limit (QNL) provided that the pump source is at the QNL~\cite{walls.book}. It is, however, possible
to enforce squeezed state production via different means. Regularizing the pumping mechanism,
e.g. using a sub-Poissonian pump, may lead to amplitude squeezing at low frequencies
of the laser field~\cite{yamamoto86.pra}. This has been shown experimentally for semiconductor lasers, pumped
with a sub-Poissonian electrical current~\cite{machida87.prl}. Alternatively, squeezed states
can be produced by some mechanisms intrinsic to the laser medium. For example
if the decay rate from the lower lasing level and the pumping rate are matched~\cite{khazanov90.pra}
or if the coherence effect between
the pump levels plays a significant role~\cite{ralph91.pra}, amplitude squeezing is expected to occur.
Finally, squeezing effects are anticipated to appear if a nonlinear $\chi^{(2)}$-crystal is
placed inside the laser cavity~\cite{gorbachev89.zet,fernandez89.QO,walls90.pra,schack91.pra,levien93.pra,white96.josab,zhang2000.josab}.
In particular, the system comprising a polarization
degenerate (type I phase matched) SHG crystal internal to a laser has been
investigated theoretically. Garcia-Fernandez \emph{et al.}~\cite{fernandez89.QO} and Levien \emph{et al.}~\cite{levien93.pra}
showed that perfect quadrature squeezing may be obtained in the laser field and the upconverted second
harmonic field, respectively, for certain settings of the decay rates involved.
A rigorous treatment of the same system was carried out by White \emph{et al.}~\cite{white96.josab,white97.jmo}.
Contrary to other theoretical investigations, these authors
included the effect of very fast dephasing of laser
coherence, a practical fact in all common laser systems. They found that the inclusion of this effect
leads to additional noise, obscuring the squeezing to the extent that the experiment is
not feasible for squeezing production using a type I SHG crystal. This conclusion has been
further corroborated by the fact that squeezing effects from a laser with internal frequency
doubling have so far never been observed experimentally.

However, in this paper we show that one should not give up on a up-conversion process internal to
a laser as
a source of efficiently squeezed light. Rather than incorporating a type I crystal inside the
laser cavity, as done in ref.~\cite{white96.josab}, we here consider the case of a type II
phase matched crystal placed inside
the cavity. In a type II SHG process, two fundamental beams along the
crystal axes have to be excited. We show that this extra degree of polarization freedom
may give rise to very efficient squeezing of a certain polarization state of the fundamental field.

The way of attacking the problem is to derive a
set of linearized Langevin equation for the fluctuations of the atomic variables and the fundamental
fields (section~\ref{sec2}) and together with the steady state values derive
the spectrum for the relevant quadrature (section~\ref{sec3}). In section~\ref{sec3} we
discuss the experimental feasibility and calculate the degree of squeezing that might be expected in
a realistic system. Finally, in section~\ref{sec4} we conclude this work.

\section{Linearized quantum Langevin equations}
\label{sec2}

Many different approaches have been developed to characterize the quantum fluctuations
of a laser.
Here we use a linearized
input-output approach in which linearized quantum Langevin equations, derived from the Hamiltonian,
are solved directly in Fourier space to yield solutions for the quadrature noise spectra.
Details regarding such derivation for a laser system
(without a nonlinear crystal) can be found in ref.~\cite{ralph96.pra}.
In this section we use this approach to describe a system consisting of a laser crystal
and a type II nonlinear crystal for SHG located inside a commen cavity.

\begin{figure}[h] \centering \includegraphics{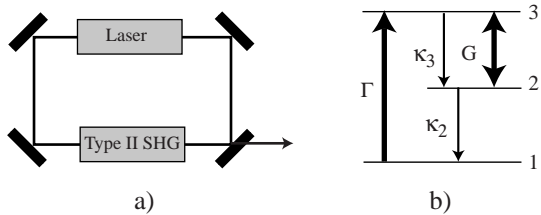} \caption{\it a) Schematic setup
of the self-frequency-doubling laser. b) Energy level scheme of the lasing atoms.}
\label{app2_laser}  \end{figure}

\begin{figure}[h] \centering \includegraphics{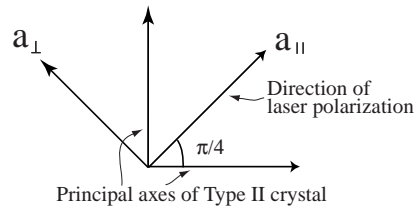} \caption{\it The principal axes of the
nonlinear crystal with respect to the base in which the theory is derived.}
\label{bases}  \end{figure}

A schematic diagram of the model is shown in Fig.~\ref{app2_laser}a.
The frequency doubling laser consists of a four level laser crystal
(with one of the levels adiabattically eliminated),
a type II nonlinear medium and two orthogonally polarized ring modes;
the parallel polarized mode, $a_\|$, and
the orthogonally polarized mode, $a_\bot$ (see Fig.~\ref{bases}). The former mode
constitutes the laser mode as well as one of the
fundamental modes of the type II SHG process. This mode is
therefore coupled to the lasing atoms via the
coupling strength, $g$, and to the nonlinear crystal via the coupling parameter, $\kappa$.
The orthogonally polarized mode also has to be taken into account due to the
nondegenerate polarization character of the SHG process. We assume that this mode does not
couple directly to the laser medium. This is a good assumption as long as the mode occupies the
vacuum state, but it becomes invalid when the mode is intense. However, the implementation of an
extra ring cavity, which is coupled to the laser cavity around the SHG-crystal,
assures that this assumption is valid. We should point out that we consider the system
in a basis which is rotated 45$^o$ with respect to the basis spanned by the principal axes
of the nonlinear crystal~\cite{ou94.pra} as shown in Fig.~\ref{bases}.
The reason for this basis shift is to make
the intense laser field parallel to one of the fundamental fields and also
because, as we will see soon, these polarization directions are particularly
interesting.

The Hamiltonian describing the internal mode interaction in the laser- and the SHG-process is
\begin{eqnarray}
H&=&i\hbar g \left(a_{\|}^\dagger \sigma_{23}-a_{\|}\sigma_{23}^\dagger\right)
+i\frac{\hbar\kappa}{2}\left(ba_\|^{\dagger\; 2}-ba_\bot^{\dagger\; 2}-h.c.\right)
\label{hamil}
\end{eqnarray}
where $\sigma_{23}$ and
$\sigma_{23}^\dagger$ are the collective
atomic lowering and raising operators between level 2 and 3.
The second harmonic mode is represented by the annihilation operator $b$ and the
creation operator $b^\dagger$.
Using the interaction Hamiltonian (\ref{hamil}) together
with the reservoir Hamiltonian a set of quantum Langevin equations of motion
for the internal field operators, the occupation operators and the coherence operators
can be derived directly~\cite{ralph96.pra}.

We introduce a series of
assumptions consistent with most laser systems with intra-cavity SHG. Firstly, we
assume that the decay rate of the second harmonic field is high; the field escapes the cavity
immediately after its generation, enabling an adiabatical elimination from the Langevin equations.
Furthermore, we assume that the laser coherence, the pump coherence, the pump cavity and
the upper level decay very rapidly. This enables an adiabatical elimination of the pump mode and
the occupation operator of the upper level (see ref.~\cite{ralph96.pra} for details). This results
in a three-level laser description as shown in Fig. \ref{app2_laser}b.

The steady-state solutions for the atomic populations and the fields are required to
evaluate the field spectra. The semiclassical equations of motion are given by
\begin{eqnarray}
\frac{d\langle a_\|\rangle}{dt}&=&\frac{G}{2}(\langle \sigma_3\rangle-\langle \sigma_2\rangle)\langle a_\|\rangle-\gamma_\|\langle a_\|\rangle-\mu\langle a_\|\rangle^*(\langle a_\|\rangle^2-\langle a_\bot\rangle^2)\nonumber\\
\frac{d\langle a_\bot\rangle}{dt}&=&-\gamma_\bot\langle a_\bot\rangle+\mu\langle a_\bot\rangle^*(\langle a_\|\rangle^2-\langle a_\bot\rangle^2)\nonumber\\
\frac{d\langle \sigma_1\rangle}{dt}&=&\kappa_2\langle \sigma_2\rangle-\Gamma \langle \sigma_1\rangle\\
\frac{d\langle \sigma_2\rangle}{dt}&=&G(\langle \sigma_3\rangle-\langle \sigma_2\rangle)\langle a_\|\rangle^2+\kappa_3\langle \sigma_3\rangle-\kappa_2\langle \sigma_2\rangle\nonumber\\
\frac{d\langle \sigma_3\rangle}{dt}&=&-G(\langle \sigma_3\rangle-\langle \sigma_2\rangle)\langle a_\|\rangle^2-\kappa_3\langle \sigma_3\rangle+\Gamma \langle \sigma_1\rangle\nonumber
\label{app_classic}
\end{eqnarray}
$\langle\sigma_i\rangle$ is the population of level $i$ scaled by the number of atoms $N$
and $\langle a_n\rangle$ $(n=\bot ,\| )$
are the amplitudes scaled by $\sqrt{N}$.
The total decay rates for the ring modes are $\gamma_\|$ and $\gamma_\bot$, the pump rate is
denoted by $\Gamma$ and the decay rate between level 3 and 2 (2 and 1)
is given by $\kappa_3$($\kappa_2$). $G$
is the stimulated emission rate per photon for the laser mode and
the non-linear coupling parameter is $\mu$, which is proportional to $\kappa^2$.

\begin{figure}[hbt] \centering \includegraphics[width=7cm]{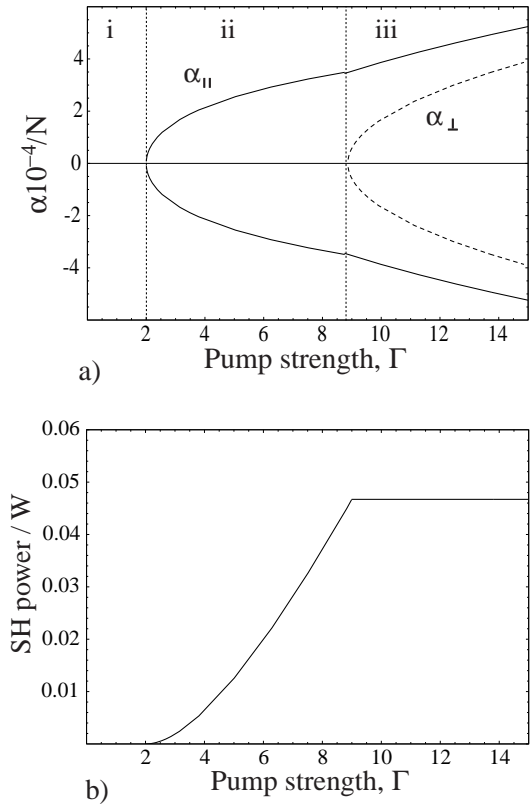} \caption{\it
Steady-state solutions as a function of the pump strength, $\Gamma$. Solid curve in (a) represents the
parallel polarized field (the laser field) amplitude while the
dashed curve is the orthogonally polarized mode.
(b) the second harmonic power versus the coupling strength.}
\label{app2_laser2}  \end{figure}

Homogeneous steady-state solutions are derived by equating (\ref{app_classic}) to zero whereby
three different solutions are found.
The solutions for the internal fields depend on the power
level of the beam, pumping the laser process, and can be divided into
three regimes as shown in Fig.~\ref{app2_laser2}a.
Below threshold for laser action,
we naturally have $\langle a_\bot\rangle=\langle a_\|\rangle=0$ (region (i)).
Above laser threshold one of the field solutions destabilizes associated with the occurrence of the parallel
polarized fundamental field, e.g. the laser field (region(ii)). Another instability, which corresponds to the
emergence of the orthogonally polarized fundamental mode, can be reached by increasing the
pump power further (region(iii)).

For completeness, the
second harmonic power is shown in Fig.~\ref{app2_laser2}b, and we observe that, due to the
back-conversion (associated with the generation of a bright orthogonally polarized mode),
the second harmonic output clamps at a certain value.

To solve for the quantum dynamics we now conduct a linearization approximation
by inserting the superpositions
\begin{equation}
w=\langle w\rangle+\delta w,\;\; w=(a_\bot,a_\|,\sigma_1,\sigma_2,\sigma_3)
\end{equation}
into the quantum Langevin equations. We then arrive
at the following set of linearized Langevin equations
for the quadrature fluctuations of the fundamental fields
\begin{eqnarray}
\frac{d\delta Z_{\|}}{dt}&=&\Theta G(\delta\sigma_3-\delta\sigma_2)\langle a_\|\rangle+2\sqrt{\mu}\langle a_\|\rangle\delta Z_b^{in}\nonumber\\
&&-2\mu \langle a_{n}\rangle^2\delta Z_{\|}\pm 2\mu\langle a_\|\rangle\langle a_\bot\rangle\delta Z_{\bot}\nonumber\\
&&+\sqrt{2\gamma_\|^l}\delta Z_{\|}^{in1}+\sqrt{2\gamma_\|^c}\delta Z_{\|}^{in2}+\sqrt{G}\delta Z_p\nonumber\\
\frac{d\delta Z_{\bot}}{dt}&=&-\gamma\delta Z_{\bot}-2\sqrt{\mu}\langle a_\bot\rangle\delta Z_{b}^{in}\\
&&+\left(\pm\mu(\langle a_\|\rangle^2-\langle a_\bot\rangle^2)-2\mu\langle a_\bot\rangle^2\right)\delta Z_{\bot} \nonumber\\
&&+ 2\mu\langle a_\bot\rangle\langle a_\|\rangle\delta Z_{\|}+\sqrt{2\gamma_\bot^l}\delta Z_{\bot}^{in1}+\sqrt{\gamma_\bot^c}\delta Z_{\bot}^{in2}\nonumber
\end{eqnarray}
where the quadrature fluctuation amplitudes, $\delta Z=(\delta X,\delta Y)$, of the fields
are defined by $\delta a=\frac{1}{2}(\delta X-i\delta Y)$, where $\delta X$ and $\delta Y$
represent the amplitude and phase quadratures, respectively. $n=\|$ and
$n=\bot$ apply to the amplitude and the phase quadrature,
respectively. $\Theta =1$ for the amplitude quadrature while
$\Theta =0$ for the phase quadrature. In both
equations the `plus' sign and `minus' sign are associated with the amplitude and phase, respectively.
The operators $\delta Z^{in}$ represent the input fluctuations associated with the
various passive loss mechanism
while $\delta Z_p$ is the fluctuations in the laser pump mode.
$\delta\sigma_2$ and $\delta\sigma_3$ are fluctuations associated with the
atomic populations in the two lasing levels; level 2 and 3. Finally, the decay rates for the
two fundamental fields, $\gamma_n^a$
and $\gamma_n^l$ ($n=\bot ,\|$), are associated with the losses for the cavity mirrors
and the other passive losses, respectively.

\section{Squeezing result}
\label{sec3}

From the Langevin expressions we see that the quadratures for the parallel polarized mode
are, not surprisingly, directly coupled to the laser process.
This means that the laser noise adds considerable noise to the parallel polarized mode, which is
subsequently transferred into the orthogonally polarized mode. Therefore, we might expect
that the laser source has an adverse impact on the production of squeezed light, an expectation
also alluded to by White \emph{et al.}~\cite{white96.josab} for type I SHG inside a laser cavity.
However, we see that in the regime
where the orthogonally polarized mode is unexcited, corresponding to region (ii) in
Fig.~\ref{app2_laser2}a, i.e. $\langle a_\bot\rangle=0$, the mode decouples
from the parallel polarized mode. In turn this means that the laser noise stays in the latter
mode and the orthogonal polarization mode evolves independent of the laser mode.
Setting $\langle a_\bot\rangle=0$ in the Langevin equation for the phase quadrature for the orthogonally
polarized mode we find
\begin{equation}
\frac{d\delta Y_{\bot}}{dt}=-\gamma_\bot\delta Y_{\bot}
-\mu\langle a_\|\rangle^2\delta Y_{\bot}+\sqrt{2\gamma_\bot^l}\delta Y_\bot ^{in1}+\sqrt{2\gamma_\bot^c}\delta Y_\bot ^{in2}
\end{equation}

We proceed by evaluating the output fluctuations of the orthogonally polarized mode,
\begin{equation}
\delta Y_\bot ^{out}=\sqrt{2\gamma_\bot^c}\delta Y_\bot -\delta Y_\bot^{in2},
\end{equation}
in frequency space, and subsequently derive the phase quadrature spectrum:
\begin{equation}
V_\bot(\omega)=1-\frac{2\gamma_\bot^c\left(G\left(\langle \sigma_3\rangle-\langle \sigma_2\rangle\right)-2\gamma_\|\right)}{\left(\gamma_\bot-\gamma_\|+\frac{1}{2}G\left(\langle \sigma_3\rangle-\langle \sigma_2\rangle\right)\right)^2+\omega^2}
\label{variance}
\end{equation}
where $\omega$ is the analyzing frequency and
\begin{eqnarray}
\langle \sigma_3\rangle&=&\frac{1}{2a}\left(-b+\sqrt{b^2-4ac}\right)\nonumber\\
\langle \sigma_2\rangle&=&\Gamma\frac{1-\langle \sigma_3\rangle}{\kappa_2+\Gamma}\nonumber\\
a&=&\frac{G^2}{2\mu}\left(1+\frac{\Gamma}{\Gamma+\kappa_2}\right)^2\nonumber\\
b&=&\kappa_3+\kappa_2\frac{\Gamma}{\Gamma+\kappa_2}\nonumber\\
&&-G\left(1+\frac{\Gamma}{\Gamma+\kappa_2}\right)\left(\frac{G}{\mu}\frac{\Gamma}{\Gamma+\kappa_2}+\frac{\gamma_\|}{\mu}\right)\nonumber\\
c&=&G\frac{\Gamma}{\Gamma+\kappa_2}\left(\frac{G}{2\mu}\frac{\Gamma}{\Gamma+\kappa_2}+\frac{\gamma_\|}{\mu}\right)-\kappa_2\frac{\Gamma}{\Gamma+\kappa_2}\nonumber\\
\end{eqnarray}
This corresponds to the steady-state solution indicated by
region (ii) in Fig.~\ref{app2_laser2}a.

We immediately see that the expression for the spectrum (\ref{variance}) resembles the one for a
subthreshold optical parametric oscillator~\cite{drummond81.oa}.
It is a well known fact that such a process is capable of producing perfect squeezing near
its oscillation threshold~\cite{collett85.pra}.

Strictly speaking the linearization approximation
breaks down when the steady-state solution becomes comparable to the size of its fluctuations.
Therefore, in the above mentioned case where $\langle a_\bot\rangle =0$,
the linearization approximation is invalid.
However, in a realistic system a small part of the parallel polarized
field will always be coupled into
the orthogonally polarized wave due to inevitable imperfect alignment of the direction of the
laser polarization. In fact, only a few microwatts have to be transferred from the laser polarization
to the orthogonal polarization mode to justify the linearization approximation. Due to the
vanishingly small amount of light, which is transferred between the two polarization
modes, the seeded light will be at the QNL. Therefore, the asymmetry will not
be detrimental to the production of squeezed light.

\section{Experimental considerations}
\label{sec4}

We now proceed with a discussion of the experimental feasibility of generating
squeezed light in the system mentioned above under realistic conditions.

A practical setup has to fulfill a number of important criteria:
It is important to maintain low losses at the orthogonally polarized
states in order to obtain high squeezing. Both the parallel and
orthogonal polarization states have to be kept simultaneously at resonance.
It is crucial to make the type-II phase matched nonlinear crystal work as
a full wave plate to avoid unwanted cross coupling between the two polarization
states. Finally beam walk-off has to be eliminated since it introduces
cross coupling between the polarization states.
A setup that fulfills the above mentioned design criteria is shown in Fig.~\ref{sqz2}.
Here a standard unidirectional Bow-tie ring laser, confined by the mirrors M1-M4,
is coupled to a secondary resonator through polarization beam splitters e.g.
Wollaston prism polarizers (WP).
Using Wollaston prisms, low losses
are obtained in both cavities and a high extinction ratio is possible.
Therefore, the parallel polarized wave circulates in the laser
cavity while the orthogonally polarized wave is confined in the coupled cavity, which is defined by the
two prisms and the mirrors M5 and M6.
The second cavity
can be locked resonantly to the laser, using normal locking techniques by
probing the coupled cavity in the counterpropagating direction
with a fraction of the laser field that leaks out of mirror M3.

\begin{figure}[hbt] \centering \includegraphics[width=7cm]{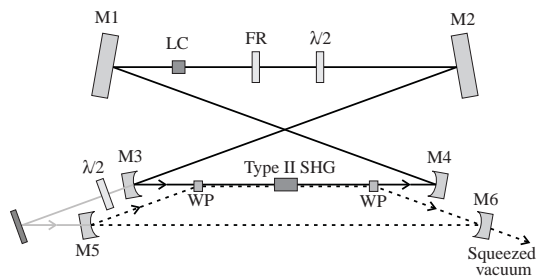} \caption{\it
Schematic of an experimental setup to realize the proposed scheme. The parallel polarized
fundamental field (the laser field) resonantes between the mirrors M1-M4, while the orthogonally polarized
field resonates in a coupled cavity confined by two Wollaston prisms (WP) and the mirrors M5 and M6.
A leakage through mirror M3 can be used for locking the coupled cavity to the desired resonance.
LC: Laser crystal, FR: Faraday rotator and $\lambda$/2: Half-wave plate.
}
\label{sqz2}  \end{figure}

A phase-matching scheme without beam walk-off has to be chosen
in order to avoid direct coupling of the fundamental fields. This is
achieved either by choosing a lasing wavelength where noncritical phasematching
can be obtained using a type II crystal e.g. KTP at 1080~nm~\cite{ou.ol}
or by co-doping the nonlinear crystal to match the desired lasing wavelength~\cite{jiyang.crystal}.
Furthermore, careful temperature control should be applied to stabilize
the non-linear crystal as a full wave plate.

Optimum squeezing in the orthogonally polarized vacuum state occurs at the point
of its oscillation, which corresponds to: $\langle a_\|\rangle^2 =\gamma_\bot/\mu$.
Thus we can estimate the diode pump strength needed to reach the point of instability and
hence optimum squeezing. Using realistic values of $\mu=8.0\cdot 10^{-4}$ s$^{-1}$ and $\gamma_\bot=1.6\cdot 10^7$ s$^{-1}$
we find $\langle a_\|\rangle^2=19.7\cdot 10^9$, which again
corresponds to a pump rate of 12.6 s$^{-1}$. This pump rate is readily reached, using
e.g. a diodepumped Nd:YVO$_4$ laser.

In Fig.~\ref{sqz2} we have plotted the expected degree of squeezing at 2~MHz as a
function of the pump rate normalized to the pump rate at the instability point.
In calculating this degree of squeezing we have used realistic parameters for the
laser medium, the cavity and the non-linear medium (see Fig. text).
Squeezing at oscillation threshold is given by
\begin{equation}
V=1-\frac{4\gamma_\bot^c\gamma_\bot}{4\gamma_\bot^2+\omega^2}
\end{equation}
From this expression and using the parameters mentioned above
we estimate that as much as 7.5~dB squeezing at 2~MHz
should be observable in a realistic experiment.

\begin{figure}[hbt] \centering \includegraphics[width=7cm]{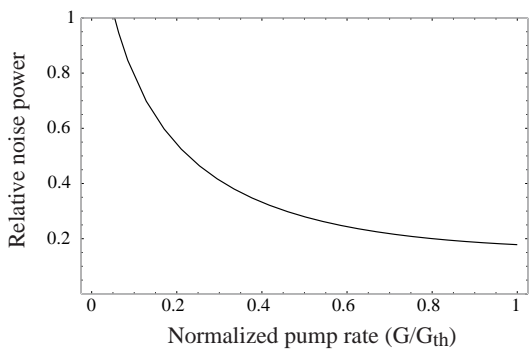} \caption{\it
The phase quadrature noise power of the orthogonally polarized fundamental field as a function of
the diode normalized pump rate. $\gamma_\bot^c=1.5\cdot 10^7$, $\gamma_\bot^l=0.75\cdot 10^6$,
$\gamma_\|^c=0.5\cdot 10^6$, $\gamma_\|^l=5.0\cdot 10^6$ and $\omega =4\pi 10^6\; s^{-1}$.}
\label{sqz2}  \end{figure}

\section{Conclusion}
\label{sec5}

We have investigated, theoretically, squeezing in a
frequency doubler based on a type II phase matched nonlinear crystal located inside
a laser cavity. The system
is resonant for two orthogonal polarizations directions of the fundamental
field (doubly resonant for the fundamental) while the generated second harmonic
field is allowed to escape freely. Contrary to previously published results
concerning type I frequency doubling, where pump laser noise couples strongly
into the quadratures of the generated laser field and subsequently into the
generated second harmonic light, we find that in the regime between the threshold
for generation of the parallel polarized fundamental laser field and the threshold
for down-conversion into the orthogonally polarized fundamental field it is
possible to obtain squeezing in the fundamental field. In fact,
we show that the Langevin equation for the phase quadrature of the orthogonally
polarized fundamental is identical to the one for sub-threshold optical parametric
oscillation. Consequently, we claim that our setup can in principle produce perfectly
squeezed light in the orthogonally polarized fundamental field near the threshold for down-conversion.
We propose an experiment for verification of this idea based on a unidirectional
bow-tie laser with a type II phase matched second order nonlinear crystal.
The resonance for the orthogonally polarized light is provided by a separate cavity,
which is aligned with the laser cavity by means of polarization dependent components.
The advantage of the proposed technique is
its simplicity compared to previous setups for subthreshold optical parametric oscillators,
which require locking between the laser cavity, the upconverter the squeezing cavity and
a reference beam.
We believe the proposed method with carefully aligned components will allow detection
of squeezed light from a frequency doubler internal to a laser, a process which has so far not
been practical due different noise sources triggered by the laser process.

\end{document}